# Gigahertz Self-referenceable Frequency Comb from a Semiconductor Disk Laser

Christian A. Zaugg,[1] Alexander Klenner,[1] Mario Mangold,[1] Aline S. Mayer,[1] Sandro M. Link,[1] Florian Emaury,[1] Matthias Golling,[1] Emilio Gini,[2] Clara J. Saraceno,[1,3] Bauke W. Tilma,[1] Ursula Keller[1]

[1]*Department of Physics, Institute for Quantum Electronics, ETH Zürich, 8093 Zürich, Switzerland*
[2]*FIRST Center for Micro- and Nanoscience, ETH Zürich, 8093 Zürich, Switzerland*
[3]*Laboratoire Temps-Fréquence, Université de Neuchâtel, 2000 Nechâtel, Switzerland*

[*]*zauggc@phys.ethz.ch*

**Abstract:** We present a 1.75-GHz self-referenceable frequency comb from a vertical external-cavity surface-emitting laser (VECSEL) passively modelocked with a semiconductor saturable absorber mirror (SESAM). The VECSEL delivers 231-fs pulses with an average power of 100 mW and is optimized for stable and reliable operation. The optical spectrum was centered around 1038 nm and nearly transform-limited with a full width half maximum (FWHM) bandwidth of 5.5 nm. The pulses were first amplified to an average power of 5.5 W using a backward-pumped Yb-doped double-clad large mode area (LMA) fiber and then compressed to 85 fs with 2.2 W of average power with a passive LMA fiber and transmission gratings. Subsequently, we launched the pulses into a highly nonlinear photonic crystal fiber (PCF) and generated a coherent octave-spanning supercontinuum (SC). We then detected the carrier-envelope offset (CEO) frequency ($f_{CEO}$) beat note using a standard $f$-to-$2f$-interferometer. The $f_{CEO}$ exhibits a signal-to-noise ratio of 17 dB in a 100-kHz resolution bandwidth and a FWHM of ≈10 MHz. To our knowledge, this is the first report on the detection of the $f_{CEO}$ from a semiconductor laser, opening the door to fully stabilized compact frequency combs based on modelocked semiconductor disk lasers.

## 1. Introduction

Passively modelocked optically pumped vertical external-cavity surface-emitting lasers (VECSELs) [1] have become very interesting ultrafast laser sources delivering high average powers in the gigahertz pulse repetition rate regime [2]. State-of-the-art modelocking with semiconductor saturable absorber mirrors (SESAMs) [3, 4] offers the full exploitation of the band-gap engineering of the semiconductor technology, resulting in the extension of the spectral coverage from modelocked VECSELs below [5, 6] and above [7, 8] the typical range of around 1 µm. Optimized for low saturation fluences [9], SESAMs fulfill a key requirement for short pulses at high powers from VECSELs [10, 11]. Furthermore, the integration of both gain and absorber section into MIXSELs (modelocked-integrated external cavity surface emitting lasers) [12] has lead to extremely compact femtosecond pulsed sources [13, 14].

The gigahertz operation regime of semiconductor disk lasers (SDL) such as VECSELs and MIXSELs makes them very promising sources for compact frequency combs with large comb-tooth spacing and high-power per mode [13], suitable for applications for example in metrology [15, 16], ultra-stable optical clocks [17] or spectroscopy [18, 19].

To date, gigahertz frequency combs are most commonly based on ultrafast bulk solid-state lasers using gain media with long upper-state lifetimes such as diode-pumped Yb-doped [20] or green-pumped Ti:Sapphire lasers [21]. Although impressive performance has been obtained with these systems, they do not offer the benefits of semiconductor based gain materials in terms of low-cost mass-production, integration and compactness. Furthermore, scaling femtosecond operation to the multi-GHz regime becomes increasingly difficult because small gain cross sections and low pulse energies introduce Q-switching instabilities [22]. Continuous progress in the performance of ultrafast SDLs makes these pulse sources very attractive for frequency comb generation. Stable fundamental modelocking in an extremely wide range of repetition rates between <100 MHz [23] and >100 GHz [13] is possible with modelocked SDLs. Record performance is achieved beyond 1 GHz [24-26], and even continuous repetition rate tuning during operation was demonstrated [13, 27, 28]. Furthermore, modelocked SDLs exhibit excellent timing jitter and amplitude noise properties [29-33] and can even be stabilized to an extremely low level [30-33]. In terms of pulse duration, 107 fs [34] in fundamentally modelocked operation or even sub-100 fs in burst operation [35] were demonstrated, however with a limited average output power of only a few miliwatts. With the first watt-level femtosecond modelocked VECSEL [24] a new operation regime was established. More recently, even 5.1 W of average output power with 682-fs pulses [25] or 3.3 W with 400-fs pulses were reported [26] with a record-high peak power of 4.35 kW.

A modelocked laser generates a frequency comb with two degrees of freedom: the spacing of the comb defined by the pulse repetition frequency $f_{rep}$ and the translation of the full comb defined by the carrier-envelope offset (CEO) frequency $f_{CEO}$. The detection and stabilization of $f_{rep}$ is well established since the 1980[th] [36, 37] and was successfully implemented with SDLs [30-33]. The $f_{CEO}$ requires more sophisticated detection schemes as initially proposed by Telle et al [15]. The most established technique is the $f$-to-$2f$-interferometer [15], which requires the generation of a coherent octave-spanning supercontinuum (SC). Such a broad spectrum is typically generated using an additional external highly nonlinear photonic crystal fiber (PCF) [38-40]. Recent progress with SESAM modelocked VECSELs look very promising because they approach the peak power required for SC generation as demonstrated before with diode-pumped solid-state lasers in the gigahertz repetition rate regime [20]. So far, however, launching the modelocked VECSEL output directly into a highly nonlinear PCF did not generate a sufficiently broad continuum [26]. Even additional passive compression to 235-fs pulses with 280 mW average output power did not enable the generation of an octave-spanning SC [41] because not enough average power could be delivered to the PCF.

A straightforward solution for providing more average power is external amplification, e.g. with rare earth doped fiber amplifier in a master oscillator power amplifier (MOPA) configuration. Recently, multi-stage Yb-doped fiber amplifier systems were implemented for modelocked VECSELs, resulting in several tens of Watts of average output power [42, 43] and even in SC generation covering several hundreds of nanometers [43, 44]. But despite the high power, the coherent octave-spanning SC and the $f_{CEO}$-detection were not yet obtained, most likely because pulse durations well below 200 fs are crucial in order to maintain the coherence of a SC generated in a PCF. This has been numerically investigated and experimentally verified with a gigahertz DPSSL [20, 45].

Here, we present the first measurement of the $f_{CEO}$ of a modelocked VECSEL to the best of our knowledge. In order to reach sufficient peak power for the detection, we use a single-pass fiber amplification stage and subsequent compression. The SESAM modelocked VECSEL delivered 100 mW average power in 231 fs pulses at a repetition rate of 1.75 GHz. We used 5.5 W for the compression and obtained 85 fs pulses with 2.2 W average power and >10 kW peak power, which was sufficient to generate a coherent octave-spanning SC in a commercially available highly-nonlinear PCF. Launching this SC into a standard $f$-to-$2f$-interferometer allowed us to detect the $f_{CEO}$ exhibiting a signal-to-noise ratio (SNR) of ≈17 dB (RBW = 100 kHz) and a 3 dB bandwidth of ≈10 MHz.

This first proof-of-principle experiment demonstrates the potential for frequency comb generation with ultrafast VECSELs or even MIXSELs. We believe further performance improvements will make external amplification and maybe even compression obsolete in the near future.

## 2. Experimental setup

### 2.1 SESAM modelocked VECSEL

Both the SESAM and VECSEL chip were grown and processed in the FIRST Center for Micro- and Nanoscience at ETH Zürich. We use a VECSEL gain chip that is optimized for low group delay dispersion (GDD) around the lasing wavelength and high gain saturation fluence, suitable for short pulse generation according to the previously published guidelines [10, 11]. The structure was grown using metal-organic vapor deposition epitaxy (MOVPE) and consists of the following elements (from the bottom to top): A 24-pair GaAs/AlAs distributed Bragg reflector (DBR) designed for a wavelength of 1030 nm; the active region consisting of 5 pairs of $GaAs_{0.94}P_{0.06}$ strain compensated InGaAs quantum wells (QW) separated by 26 nm, each pair centered around an anti-node of the standing electric field pattern; a hybrid semiconductor-dielectric low-GDD anti-reflection coating with 4 pairs of $AlAs/Al_{20}Ga_{80}As$, a layer of GaAs and a layer of fused silica (FS). The layer-stack (except the FS layer) was grown in reverse order and subsequently In-soldered on a chemical vapor deposition (CVD) grown diamond heat spreader according to the flip-chip bonding and wet-chemical etching scheme described in [46]. The final layer of FS was deposited on the chip using plasma enhanced chemical vapor deposition (PECVD).

For passive modelocking, a strained low-temperature QW saturable absorber SESAM was grown by molecular beam epitaxy (MBE) identical to the ones characterized in [14, 47], but adapted to 1030 nm. The QW-SESAM consists of a 30-pair GaAs/AlAs DBR and a single InGaAs QW embedded in AlAs, arranged in the anti-node of the standing electric field pattern. The as-grown structure is finished such that minimal electric field strength is present at the semiconductor-air interface. Subsequently, a $\lambda/4$-layer of SiNx was deposited using PECVD in order to enhance the modulation depth $\Delta R$ of the SESAM for optimum modelocking performance. The maximum absorption at room temperature occurs at 1030 nm.

For modelocking, we employ a V-shaped cavity as sketched in Fig. 1(a) with a total length of $\approx 86$ mm consisting of the gain chip as a folding mirror and the SESAM and the output coupler (OC) as end mirrors. The OC is partially reflective with a transmission of 1.5% and has a radius of curvature (ROC) of 100 mm. The distance between the VECSEL and OC and the SESAM is $\approx 30$ mm and $\approx 56$ mm, respectively. The resonator mode $1/e^2$ radius is $\approx 100$ μm on the SESAM and $\approx 210$ μm on the VECSEL. The gain chip is pumped with $\approx 11$ W of power under an angle of 45°. The beam of a commercial fiber coupled (NA 0.22) 808 nm low-brightness pump ($M^2 \approx 72$) was delivered onto the VECSEL gain chip ($\approx 250$ μm radius) using standard optics. An optical isolator is used at the VECSEL output to prevent modelocking instabilities induced by potential back-reflections.

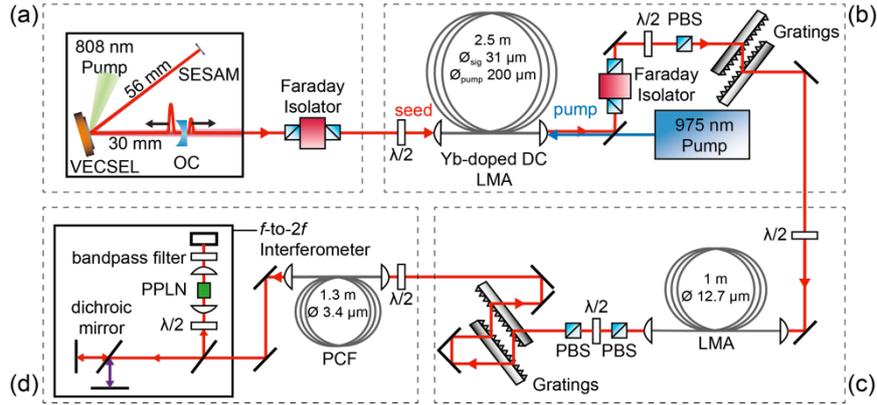

Fig. 1. Schematic of the experimental setup. (a) SESAM modelocked VECSEL; OC: output coupler; total cavity length: 86 mm; for details see section 2.1. (b) Backward-pumped fiber amplifier system, attenuation and grating compressor; DC: double-clad; LMA: large mode area; PBS: polarizing beam splitter; details in section 2.2. (c) Passive compression stage with LMA for spectral broadening and grating compressor; PBS and $\lambda$/2 plate for attenuation; see section 2.3 for details. (d) Supercontinuum (SC) generation and *f*-to-2*f*-interferometer to detect carrier envelope offset frequency; PCF: photonic crystal fiber; PPLN: periodically poled lithium niobate; details in section 2.4.

## 2.2. Fiber amplifier

The output of the modelocked VECSEL is used to seed a fiber amplifier as shown in Fig. 1(b). The Yb-doped polarization-maintaining photonic crystal fiber is commercially available (NKT DC-200/40-PZ-Yb) and suitable for low-brightness pumping. The double-clad (DC) fiber layout allows for efficient pumping via the inner cladding (diameter of 200 µm) and easy signal coupling due the LMA (mode field diameter of 31 µm). Furthermore, both fiber ends are hermetically sealed and mounted on high-power connectors with an angle-polished facet. We use a 2.5-m fiber length in order to provide sufficient gain. A standard 975 nm fiber-coupled diode laser delivering up to 160 W of average power is used to pump the fiber amplifier in a backward-pumped configuration. With a water chiller, the temperature of the pump laser heat sink is kept at ≈35°C, optimized for efficient pump-absorption in the fiber. To separate the amplified signal from the pump at the fiber output, we use an optical low-pass filter, highly reflective for the pulses centered around 1040 nm and transparent for the pump. In order to protect the system from potentially occurring back-reflections (e.g. from non-angle polished fibers), we placed a Faraday isolator at the output of the amplifier. Subsequently, a $\lambda$/2 plate and a polarizing beam splitter (PBS) were used to enable a continuous attenuation of the signal as the experiments required a high dynamic range of the average output power for a given amplification factor. In order to compensate for pulse stretching due to dispersion and self-phase modulation (SPM) accumulated by the propagation through the 2.5 m amplifier PCF, a pair of transmission gratings (single pass) with 1250 l/mm was used to re-compress the pulses to approximately the original pulse duration of the VECSEL.

## 2.3. Passive pulse compression

With our current PCFs generating an octave-spanning SC with a high temporal coherence requires ≈100 fs pulses with high peak power [45]. To fulfill these strong requirements we implemented a passive compression stage in addition to the power amplifier. The schematic is shown in Fig. 1(c). We launch the amplified and re-compressed pulses into a LMA fiber (mode field diameter of 12.7 µm and a length of 1 m) in order to generate additional spectral components via SPM. The commercially available fiber (NKT Photonics LMA-PM-15) is hermetically sealed on both ends and terminated with high-power angled physical contact (APC) connectors. The average power after the LMA is controlled with an attenuation stage

(PBS, $\lambda/2$ wave plate, PBS). In the next step, the pulses were re-compressed with a double pass through a second pair of transmission gratings (1250 l/mm).

*2.4. SC generation and $f_{CEO}$ detection*

A PCF and *f*-to-2*f* interferometer as shown in Fig. 1(d) were finally used to generate a coherent octave-spanning SC and detect the $f_{CEO}$ beat signals. The amplified and compressed pulse train was launched into a highly nonlinear polarization maintaining PCF with a core diameter of 3.2 µm and a length of ≈1.3 m (NKT Photonics NL-3.2-945) with the same properties as the one used in [20]. The fiber has a zero-dispersion wavelength of ≈945 nm, a nonlinear coefficient of $\gamma$≈23 W$^{-1}$km$^{-1}$ and provides anomalous dispersion at 1042 nm of $\beta_2$≈-15.4 ps$^2$km$^{-1}$. The fiber input is hermetically sealed and equipped with an FC/PC connector, whereas the output has a bare end to enable easy adjustment of the fiber length. The SC was sent into a quasi-common-path *f*-to-2*f*-interferometer [15, 20, 48] to detect the $f_{CEO}$ beats of the modelocked VECSEL. In the Michelson-type interferometer a dichroic mirror separates the long- (1360 nm) and short- (680 nm) wavelength parts of the SC. The notch-type NIR-mirror transmits residual 1030-nm spectral components. After matching the temporal delay between the 1360-nm and 680-nm components, both beams are recombined. The long wavelength beam is frequency-doubled in the periodically poled lithium niobate (PPLN) crystal heated to 45°C. After a band-pass filter (680-nm center-wavelength, 10-nm FWHM ), both red beams interfere on the photodetector, generating the $f_{CEO}$ beat note. The signal from the high-speed, high-sensitivity avalanche photodetector (Menlo Systems, APD210, $f_{3dB}$ = 800 MHz) is amplified by 25 dB and then analyzed with a microwave spectrum analyzer (MSA).

## 3. Results

*3.1 SESAM modelocked VECSEL emitting 231 fs pulses*

The heat sink temperatures of both the VECSEL and SESAM were stabilized to 36.5°C and 19.2°C, respectively, using water-cooled Peltier elements. We optimized the laser for short, clean and pedestal-free pulses and stable reliable operation. A full characterization of the pulse train with an average output power of 100 mW (measured after the isolator) is presented in Fig. 2. The spectrum of the laser is centered around ≈1038 nm and covers a full width half maximum (FWHM) of $\Delta\lambda$=5.5 nm, as shown on a linear and logarithmic scale in Fig. 2(a) and (b), respectively. The pulse duration of 231 fs (Fig. 2c), is measured using a second harmonic autocorrelator. The time-bandwidth product of 0.355 indicates almost transform-limited sech$^2$-pulses. With a 50-GHz InGaAs photodetector and a MSA, we measured a pulse repetition rate of 1.75 GHz. A full span measurement (from 0 up to 13 GHz) with a resolution bandwidth (RBW) of 30 kHz is presented in Fig. 2(d), showing several equally powerful harmonics with a side-band suppression of more than 30 dB. The inset shows a zoom (span: 500 kHz; RBW: 100 Hz) into the repetition rate with a signal-to-noise ratio of >50 dB. To confirm an ideal Gaussian beam, we measured the beam quality of our laser and found the value of M$^2$=1. The laser was running stably and reliably for several hours a day over several weeks.

A key requirement for the generation of such short pulses with VECSELs is the dispersion management of the gain chip [10]. In Fig. 2(e), the group delay dispersion (GDD) of the unpumped VECSEL chip at room temperature is shown, measured with a setup based on white light interferometry as described in [49]. The absolute value of the GDD within the full lasing spectrum is <100 fs$^2$, limited by the measurement precision.

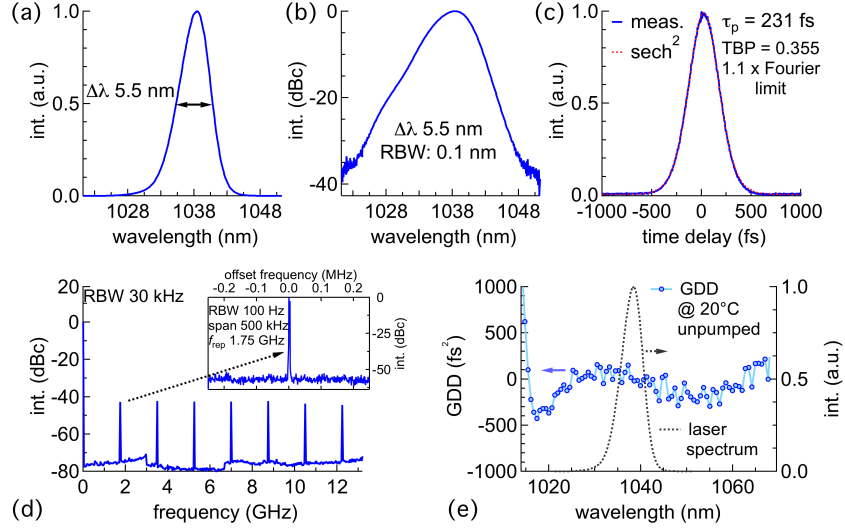

Fig. 2. Characterization of the SESAM modelocked VECSEL with 100 mW average output power. (a) Optical spectrum on a linear and (b) logarithmic scale. (c) Autocorrelation trace and sech$^2$-fit corresponding to a pulse duration of 231 fs; TBP: time-bandwidth product. (d) wide span microwave spectrum of the pulse train and zoom into the repetition rate peak (inset); RBW: resolution bandwidth. (e) Group delay dispersion (GDD) of the unpumped VECSEL gain chip measured at room temperature (blue dots) and laser spectrum (dashed line)

## 3.2. Fiber amplifier and passive compression stage

The maximum available seed power of 100 mW was coupled into the fiber amplifier. We pumped the fiber from the back through a dichroic mirror with up to 33 W resulting in an amplified signal of up to 19 W. This corresponds to an amplification factor of 22.8 dB. Neither gain saturation nor seed destabilization was observed at this pump power, indicating that even more power could be achieved from this system. We did not further amplify the pulses as only a fraction of the available power is required, and larger amplification factors could potentially increase the noise level of our pulses. In order to fully benefit from the available pump power, higher seed powers would be beneficial. A power slope and the corresponding optical-to-optical efficiency are presented in Fig. 3. The highest optical-to-optical efficiency of 57% coincides with the maximum amplification leading to 19 W of output power.

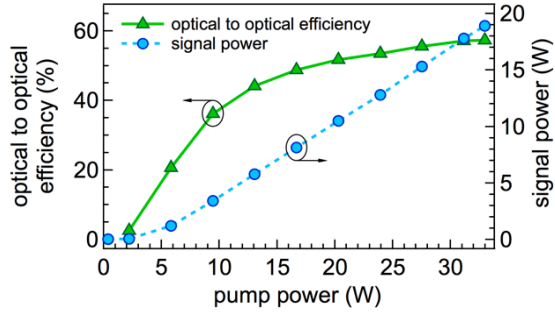

Fig. 3. Optical to optical efficiency (green triangles, left axis) and amplified signal power (blue circles, right axis) as a function of the delivered pump power using a seed average power of 100 mW.

At the maximum available power, the pulses were slightly spectrally broadened after the amplification in the fiber due to SPM. We achieved pulse durations as short as 180 fs by recompressing ≈3 W from the total of 19 W of amplified power. Although we achieved an octave-spanning SC by launching this power directly into the highly nonlinear PCF (Fig. 1(d)), no $f_{CEO}$ beat note signal could be identified, indicating an incoherent SC. We believe that the rather long pulses caused the collapse of the coherence in the PCF [20]. Numerical simulation of the SC generation using our experimental values confirm this observation [20, 45]: a pulse duration of 180 fs is at the very edge of the requirements to generate a coherent octave-spanning SC with the available highly nonlinear PCF and to detect the $f_{CEO}$ with the setup described in section 2.4. In order to achieve sufficiently short pulses, an additional passive compression stage was used.

For the further experiments, we operated the fiber amplifier at a moderate amplification factor of ≈17 dB (5.5 W directly at the amplifier output). With the gratings separated by ≈9 mm at the amplifier output, we compensate for the dispersion from the amplification fiber. We estimate the GDD introduced by the gratings to ≈-0.18±0.02 $ps^2$. At the LMA input, up to 4 W average power was available in pulses of 227 fs (Fig. 4d) and an optical spectrum with a FWHM of 5.8 nm (Fig. 4e). For more convenience, the experimental setup (Fig. 4a), the seed autocorrelation trace (Fig. 4b) and spectrum (Fig. 4c) are shown again.

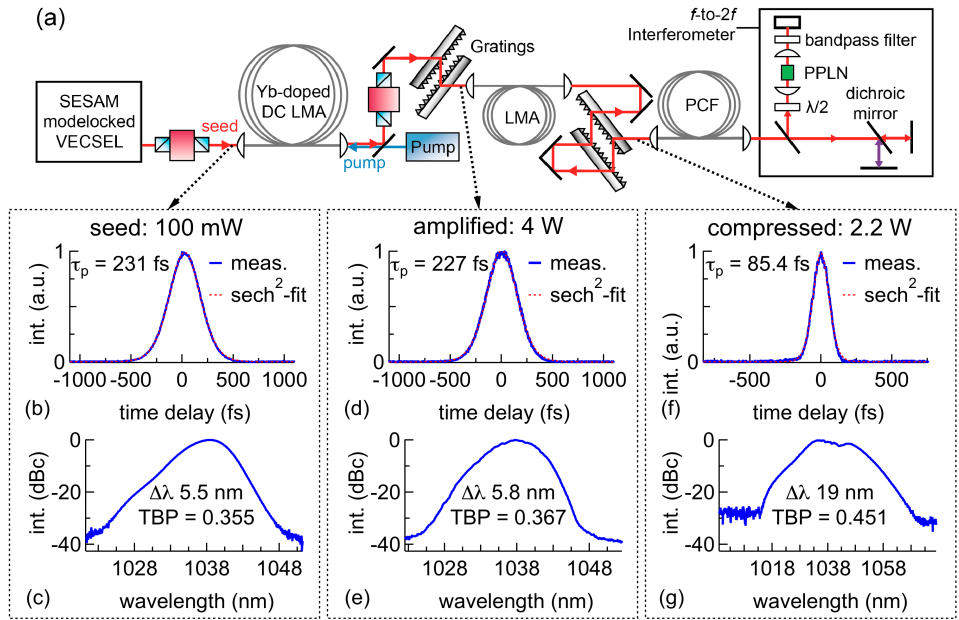

Fig. 4. Spectral and temporal pulse characterization at different stages of the experimental setup shown in (a). SESAM modelocked VECSEL seed at 100 mW: autocorrelation (b) and spectrum (c). Pulses at the output of the amplifier, after transmission through gratings at 4 W: autocorrelation (d) and optical spectrum (e). Passively compressed pulses, subsequently used for SC generation, at 2.2 W: autocorrelation (f) and spectrum (g).

We launched ≈3.8 W into the LMA fiber (mode field diameter 12.7 μm, fiber length of 1 m) with a coupling efficiency of >70%. The spectrum was broadened by SPM to 19 nm FWHM, and the pulses recompressed using transmission gratings to a pulse duration of 85 fs. We estimate the total dispersion, introduced by these double-pass gratings to ≈-0.08±0.01 $ps^2$. The very clean recompressed pulses are shown in the autocorrelation trace in Fig. 4(f). The corresponding optical spectrum is shown in Fig. 4(g). The maximum compressed power available at the input of the PCF is ≈2.2 W.

*3.4. Coherent octave-spanning SC and $f_{CEO}$ detection*

We only launched about 800 mW of the available power with the 85-fs pulses into the highly nonlinear PCF to generate an octave-spanning SC covering a spectrum from <680 nm to >1360 nm. The optical spectrum analyzer signal and the background are shown in Fig. 5(a). The dispersive wave centered at 680 nm and the Raman soliton centered at 1360 nm have a relative power of –5 dBc normalized to the 1031-nm center wavelength, sufficiently strong for stable $f_{CEO}$ beat detection.

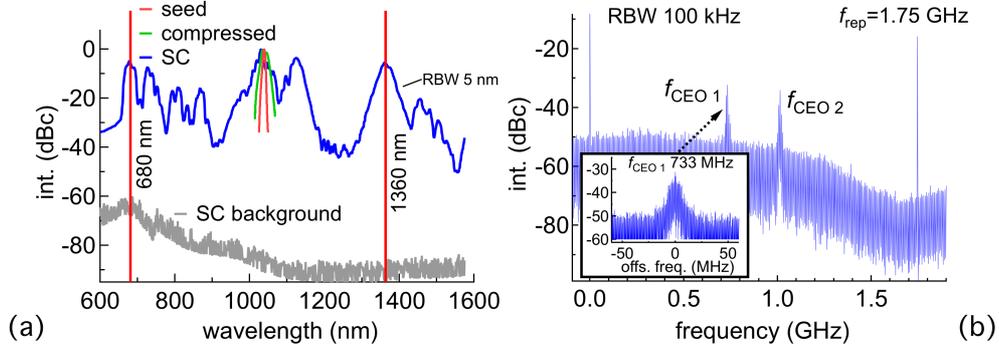

Fig. 5. (a) Coherent octave-spanning supercontinuum (blue) using the highly nonlinear PCF and spectrum analyzer background (grey). The 1360 nm and 680 nm spectral components are used for $f_{CEO}$ detection in the *f*-to-2*f*-interferometer. (b) Carrier envelope offset frequency ($f_{CEO}$) detection from the SESAM modelocked VECSEL. $f_{CEO\,1}$ and $f_{CEO\,2}$: beat notes at 733 MHz and 1017 MHz in a large span and zoom into $f_{CEO\,1}$ (inset). The decrease of both signal and noise at ≈800 MHz is due to the limited bandwidth ($f_{3dB}$≈800 MHz) of the photodiode. RBW: resolution bandwidth.

Two $f_{CEO}$ beat notes were generated by superimposing the frequency-doubled 1360-nm spectral components with the 680-nm signal on the fast photodetector (shown in Fig. 5(b) in a span of 1.9 GHz with a RBW of 100 kHz). The beat notes have a SNR of ≈17 dB and FWHM of ≈10 MHz. The inset shows a zoom into the peak located at 733 MHz. As the bandwidth of the photodiode is limited ($f_{3dB}$≈800 MHz) the signal and noise floor decrease at higher frequencies.

Besides the pure detection, a mechanism to act on the $f_{CEO}$ is a requirement for self-referencing the semiconductor frequency comb. Therefore, we investigated the influence of a modulation of the VECSEL pump current and observed a shift at a rate of ≈5 MHz/100 mA (at a total pump current of ≈22 A), i.e. ≈1% of pump current modulation shifts the beat note by its FWHM.

**4. Conclusion and outlook**

We present an important proof-of-principle experiment towards a fully stabilized frequency comb from a SESAM modelocked VECSEL by detecting the $f_{CEO}$ beat of such a laser. The SESAM modelocked VECSEL was used as a seed for this experiment followed by additional amplification and compression stages. This seed emits 231-fs pulses with an average output power of 100 mW at a repetition rate of 1.75 GHz. The pulses, centered around 1038 nm, were almost transform-limited with a FWHM of 5.5 nm. The laser was stable and reliable over several weeks.

In order to achieve sufficient peak power required for the generation of a coherent octave-spanning SC in a highly nonlinear PCF and $f_{CEO}$ detection with an *f*-to-2*f*-interferometer, we used a fiber based amplifier and passive compression scheme prior to the PCF. With that system, we obtained pulses as short as 85 fs with an average power of up to 2.2 W, which was more than sufficient for SC generation. We launched this SC into a standard *f*-to-2*f*- interferometer, and detected a beat note with a SNR of 17 dB (RBW 100 kHz) and a FWHM of ≈10 MHz. By acting on the pump laser of the VECSEL, we observed a shift of the $f_{CEO}$ of ≈5 MHz/100 mA, which should enable full stabilization of the frequency comb with state-of-the art electronics.

With the recent advances of SESAM modelocked VECSELs [25, 26], we think that further optimized gain chips will provide pulse durations in the 100 fs regime with watt-level average output powers. This will be sufficient to make the amplifier and compression stages obsolete and enable an all-passive $f_{CEO}$ detection scheme. We therefore believe that ultimately a new class of very compact and inexpensive fully stabilized gigahertz frequency combs will emerge, based on modelocked VECSELs and MIXSELs.

**Acknowledgments**

The authors acknowledge support of the technology and cleanroom facility FIRST of ETH Zürich for advanced micro- and nanotechnology. This work was financed by the Swiss Confederation Program Nano-Tera.ch, which was scientifically evaluated by the Swiss National Science Foundation (SNSF) and the ETH Research Grant ETH-26 12-1.